\documentclass{emulateapj}

\shorttitle{Reconstructing the triaxiality of Abell~1689}
\shortauthors{MORANDI, PEDERSEN, \& LIMOUSIN}

\usepackage{hyperref}

\def\aap{A\&A}

\def\apj{ApJ}

\def\mnras{MNRAS}

\def\aj{AJ}

\def\lesssim{\mathrel{\hbox{\rlap{\hbox{\lower4pt\hbox{$\sim$}}}\hbox{$<$}}}}
\def\gesssim{\mathrel{\hbox{\rlap{\hbox{\lower4pt\hbox{$\sim$}}}\hbox{$>$}}}}
\newcommand{\bu}{{\bf u}}
\newcommand{\br}{{\bf r}}
\def\lesssim{\mathrel{\hbox{\rlap{\hbox{\lower4pt\hbox{$\sim$}}}\hbox{$<$}}}}
\def\gesssim{\mathrel{\hbox{\rlap{\hbox{\lower4pt\hbox{$\sim$}}}\hbox{$>$}}}}
\newcommand{\mathbfit}[1]{\textbf{\textit{#1}}}

\begin{document}

\title{Reconstructing the triaxiality of the galaxy cluster Abell~1689: solving the X-ray and strong lensing mass discrepancy}

\author{Andrea Morandi\altaffilmark{1}, Kristian Pedersen\altaffilmark{1}, Marceau Limousin\altaffilmark{2,1}}
\altaffiltext{1}{Dark Cosmology Centre, Niels Bohr Institute, University of Copenhagen, Juliane Maries Vej 30, DK-2100 Copenhagen, Denmark}
\altaffiltext{2}{Laboratoire d'Astrophysique de Marseille, Universit\'e de Provence, CNRS, 38 rue Fr\'ed\'eric Joliot-Curie, F-13388 Marseille Cedex 13, France}


\begin{abstract}
We present the first determination of the intrinsic triaxial shapes and tree-dimensional physical parameters of both dark matter (DM) and intra-cluster medium (ICM) for the galaxy cluster Abell~1689. We exploit the novel method we recently introduced \citep{morandi2010a} in order to infer the tree-dimensional physical properties in triaxial galaxy clusters by combining jointly X-ray and strong lensing data. We find that Abell~1689 can be modeled as a triaxial galaxy cluster with DM halo axial ratios $1.24\pm 0.13$ and $2.37\pm 0.11$ on the plane of the sky and along the line of sight, respectively. We show that accounting for the three-dimensional geometry allows to solve the discrepancy between the mass determined from X-ray and strong gravitational lensing observations. We also determined the inner slope of the DM density profile $\alpha$: we measure $\alpha=0.90\pm 0.05$ by accounting explicitly for the 3D structure for this cluster, a value which is close to the cold dark matter (CDM) predictions, while the standard spherical modeling leads to the biased value $\alpha =1.16\pm 0.04$. Our findings dispel the potential inconsistencies arisen in the literature between the predictions of the CDM scenario and the observations, providing further evidences that support the CDM scenario.
\end{abstract}


\keywords{cosmology: observations – galaxies: clusters: general – galaxies: clusters: individual (Abell~1689) – gravitational lensing: strong – gravitational lensing: weak – X-rays: galaxies: clusters}



\section{Introduction}\label{intro}

Abell~1689 is a massive cluster with a redshift of $z = 0.183$ and with the largest Einstein radius, around $45$ arcsec for $z_s=1$, observed to date \citep[][]{broadhurst2005,Limousin2007}. It has been proposed as a standard example of relaxed object in hydrostatic equilibrium \citep[][]{Xue2002,lemze2010}, but the mass derived from the X-ray measurement is twice as small as that found from strong gravitational lensing at most radii \citep[][]{andersson2004,lemze2008,riemer2009,Peng2009}.

\cite{riemer2009} showed that the discrepancy between X-ray and weak lensing mass determinations has been reduced if we exclude a cool clump plus some substructure in the North Eastern part of the cluster, nevertheless a departure of factor of $\sim 2$ between SL and X-ray mass still remains in the strong lensing region.

Since lensing is sensitive to the integrated mass contrast along the line of sight, departures from the spherical assumption could justify the disagreement between X-ray and lensing mass profiles found in the literature \citep{Gavazzi2005}. In particular our recent work \citep{morandi2010a} presented the first determination of the intrinsic shapes and the physical parameters of both DM and ICM in the triaxial galaxy cluster MACS\,J1423.8+2404 by combining X-ray and lensing data, and we showed that triaxiality allows to solve the long-standing discrepancy between galaxy cluster masses determined from X-ray and gravitational lensing observations. Since cluster mass measurements are sensitive to the assumptions about symmetry, this suggests that clusters with prominent strong lensing features are not spherically symmetric and preferentially elongated along the line of sight increasing the magnitude of the lensing \citep{Oguri2009,meneghetti2010}.

The main motivation for studying Abell~1689 in large detail with the best available X-ray and lensing data is to get insight into the mass discrepancy and to understand if it can be ascribed to standard spherical geometry. In this perspective we aim at unveiling its triaxial physical properties, shape and mass distribution via implementation of the novel method described in \cite{morandi2010a} and by combining X-ray and SL data.

Hereafter we have assumed a flat $\Lambda CDM$ cosmology, with matter density parameter $\Omega_{m}=0.3$, cosmological constant density parameter $\Omega_\Lambda=0.7$, and Hubble constant $H_{0}=70 \,{\rm km/s/Mpc}$. Unless otherwise stated, we estimated the errors at the 68.3 per cent confidence level.

\section{The dataset and the analysis}\label{dataan}
Here we briefly summarize the most relevant aspects of our data reduction and analysis of Abell~1689.

\subsection{Strong lensing analysis}\label{snnen2sl}
For the strong lensing analysis we refer to the findings of \cite{Limousin2007}, who presented a reconstruction of the mass distribution of the galaxy cluster Abell~1689 using detected strong lensing features from deep ACS observations and extensive ground based spectroscopy. They presented a parametric strong lensing mass reconstruction using 34 multiply imaged systems, and they inferred two large-scale dark matter clumps, one associated with the center of the cluster and the other with a north-eastern substructure. We masked out the north-eastern sector of both the 2D projected mass map and the X-ray data, in order to avoid the contribution from this secondary substructure. We masked out also the central 30 kpc, which is affected by the mass distribution of the cD galaxy. From the lensing analysis the major clump of the cluster looks elongated with a minor-major axial ratio on the plane of the sky of $1.24\pm 0.13$ and position angle of $0.4\pm 1$ degrees. The 2D projected mass map has been rebinned into elliptical annuli, whose eccentricity, centroid and position angle is the same as that inferred from \cite{Limousin2007}. Then we calculated average values of the elliptical symmetric projected mass profile $k(R)$, $R$ being the minor radius of the 2D elliptical annuli, once we masked out the central 30 kpc, which is affected by the mass distribution of the cD galaxy. We also calculated the covariance matrix $\mathbfit{C}'$ among all the measurements of $k(R)$.

\subsection{X-ray data reduction}\label{laoa}
For the X-ray analysis we take advantage of the 2 Chandra X-ray observations (observation ID 6930 and 7289) from the NASA HEASARC archive with a total exposure time of approx. 150 ks. We summarize here the most relevant aspects of the X-ray data reduction procedure for Abell~1689. The observations have been carried out using ACIS--I CCD imaging spectrometer. We reduced these observations using the CIAO software (version 4.1.2) distributed by the {\it Chandra} X-ray Observatory Center, by considering the gain file provided within CALDB (version 4.1.3) for the data in VFAINT mode. Then we have filtered the data to include the standard events grades 0, 2, 3, 4 and 6 only, and therefore we have filtered for the Good Time Intervals (GTIs) supplied, which are contained in the {\tt flt1.fits} file. We checked for unusual background rates through the {\tt lc\_sigma\_clip}, so we removed those points falling outside $\pm 3\sigma$ from the mean value. Finally, we filtered ACIS event files on energy selecting the range 300-9500 keV and on CCDs, so as to obtain an events 2 file.

\subsection{X-ray spatial and spectral analysis}\label{sp}
We outline the methodology of spatial and spectral analysis in triaxial galaxy clusters. The general idea is to measure the gas density profile in an non-parametric way from the surface brightness recovered by a spatial analysis, and to infer the observed projected temperature profile by a spectral analysis.

The images have been extracted from the events 2 files in the energy range ($0.5-5.0$ keV), corrected by the exposure map to remove the vignetting effects, by masking out the point sources. We constructed a set of $n$ ($n=57$) elliptical annuli of minor radius $r_{m}$ around the centroid of the surface brightness and with eccentricity ${\epsilon_{b'}(r)}$ fixed to that predicted from the eccentricity ${e_{b'}(r)}$ of the DM halo on the plane of the sky from SL data (see Sect. \ref{depr}). The minor radius of each annulus has been selected out to a maximum distance $R_{\rm spat}=1139 \,{\rm kpc}$, selecting the minor radii according to the following criteria: the number of net counts of photons from the source in the (0.5-5.0 keV) band is at least 200-1000 per annulus and the signal-to-noise ratio is always larger than 2. The background counts have been estimated from regions of the same exposure, which are free from source emissions.

The spectral analysis has been performed by extracting the source spectra from $n^*$ ($n^*=9$) elliptical annuli of minor radius $r^*_{m}$ around the centroid of the surface brightness and with eccentricity equal to that predicted from the eccentricity ${\epsilon_{b'}(r)}$ of the DM halo from SL data (see above). We have selected the minor radius of each annulus out to a maximum distance $R_{\rm spec}=1089 \,{\rm kpc}$, according to the following criteria: the number of net counts of photons from the source in the band used for the spectral analysis is at least 2000 per annulus and corresponds to a fraction of the total counts always larger than 30 per cent.

All the point sources have been masked out by both visual inspection and the tool {\tt celldetect}, which provide candidate point sources. Then we have calculated the redistribution matrix files (RMF) and the ancillary response files (ARF) for each annulus.

For each of the $n^*$ annuli the spectra have been analyzed by using the XSPEC \citep[][version 11.3.2]{1996ASPC..101...17A} package, by simultaneously fitting absorbed MEKAL models multiplied by a positive absorption edge as described in \cite{vikhlinin2005} to the two observations. The fit is performed in the energy range 0.6-7 keV (0.9-7 keV for the outermost annulus only) by fixing the redshift the redshift at $z=0.183$, and the photoelectric absorption at the galactic value. We consider three free parameters in the spectral analysis for the $i$th annulus: the normalization of the thermal spectrum $K_{\rm i} \propto \int n^2_{\rm e}\, dV$, the emission-weighted temperature $T^*_{\rm proj,i}$; the metallicity $Z_{\rm i}$ retrieved by employing the solar abundance ratios from \cite{grevesse1998}. Background spectra have been extracted from regions of the same exposure, which are free from source emissions.

At last we recover the electron density $n_e = n_e(r; {\epsilon_{c'}})$ both by deprojecting the surface brightness profile and the spatially resolved spectral analysis obtaining a few tens ($n=57$) of radial measurements in ellipsoidal shells. Notice the dependency of $n_e(r; {\epsilon_{c'}})$ on the eccentricity ${\epsilon_{c'}}$ of the ICM along the line of sight to be determined \citep[for further details see Appendix~A of][]{morandi2010a}.

The global (cooling-core corrected) temperature $T_{\rm ew}$ has been estimated to be $T_{\rm ew}=8.64^{+0.13}_{-0.12}$ keV and an abundance of $0.41\pm0.03$ solar value. We classify this cluster as a intermediate cooling core source (ICC) \citep[][]{morandi2007b}: we estimated a $t_{\rm cool}\simeq 3\times 10^9$ yr. The temperature profile is very regular once we masked out the north-eastern quadrant, suggesting a relaxed dynamical state (see upper panel of Fig. \ref{entps332333}).

\subsection{Joint X-ray+lensing analysis: measuring the triaxial physical properties of ICM and DM}\label{depr}
Here we briefly summarize the major findings of \cite{morandi2010a} for the joint X-ray+lensing analysis in order to infer triaxial physical properties: for further details we refer to \cite{morandi2007a,morandi2010a}.

The lensing and the X-ray emission both depends on the properties of the DM gravitational potential well, the former being a direct probe of the 2D mass profile and the latter an indirect proxy of the 3D mass profile through the hydrostatic equilibrium equation applied on the gas temperature and density. In order to infer the model parameters of both the ICM and of the underlying DM density profile, we perform a joint analysis for strong lensing and X-ray data. In this perspective we briefly outline the methodology in order to infer physical properties in triaxial galaxy clusters: 1) we start with a generalized Navarro, Frenk \& White (gNFW) triaxial model of the DM as described in \cite{jing2002}, which is representative of the total underlying mass distribution and depends on a few parameters to be determined, namely the concentration parameter $c$, the scale radius $r_{\rm s}$ and the inner slope of the DM $\alpha$ 2) following \cite{lee2003,lee2004}, we recover the gravitational potential and surface mass profile $k$ of a dark halo with such triaxial density profile 3) we solve the hydrostatic equilibrium equation for the density of the ICM sitting in the gravitational potential well previously calculated, in order to infer a theoretical three-dimensional temperature profile $T_{\rm gas}$ in a non-parametric way 4) the joint comparison of $T_{\rm gas}$ with the observed temperature and of $k$ with the observed surface mass give us the parameters of the triaxial DM density model, and therefore all the desired physical properties of ICM and DM triaxial ellipsoids (see Fig. 1).

The work of \cite{lee2003} showed that the ICM and DM halos are well approximated by a sequence of concentric triaxial distributions with different eccentricity ratio. We define $e_{b'}$ ($\epsilon_{b'}$) and $e_{c'}$ ($\epsilon_{c'}$) as the eccentricity of DM (ICM) on the plane of the sky and along the line of sight, respectively. The iso-potential surfaces of the triaxial dark halo coincide also with the iso-density
(pressure, temperature) surfaces of the intra-cluster gas.
Notice that $\epsilon_{b'}=\epsilon_{b'}(e_{b'},u,\alpha)$ and $\epsilon_{c'}=\epsilon_{c'}(e_{c'},u,\alpha)$, with $\bu \equiv \br/r_{\rm s}$, unlike the constant $e_{b'},e_{c'}$ for the adopted dark matter halo profile. In the whole range of $u$, $\epsilon_{\sigma}/e_{\sigma}$ is less than unity ($\epsilon_{\sigma}/e_{\sigma}\sim 0.7$ at the center), i.e. the intra-cluster gas is altogether more spherical than the underlying DM halo.

In order to infer the model parameters, we construct the likelihood performing a joint analysis for SL and X-ray data, to constrain the properties of the model parameters ${\bf q}$ of both the ICM and of the underlying gNFW triaxial model of the DM, with
\begin{equation}\label{aa334}
{\bf q}=(c,r_{\rm s},\alpha,e_{c'})\ ,
\end{equation}
representing the concentration parameter, scale radius, inner slope of the DM and eccentricity of the DM along the line sight, respectively.

The method works by constructing a joint X-ray+lensing likelihood:
\begin{equation}\label{chi2wwf}
{\mathcal{L}}={\mathcal{L}}_{\rm x}\cdot {\mathcal{L}}_{\rm lens}
\end{equation}
being $\mathcal{L}_{\rm x}$ and ${\mathcal{L}}_{\rm lens}$ the likelihoods coming from the X-ray and SL data, respectively.

For ${\mathcal{L}}_{\rm x}$ holds ${\mathcal{L}}_{\rm x}\propto {\exp ( -\chi_{\rm x}^2/2)}$, with $\chi^2_{\rm x}$ equal to:
\begin{equation}\label{chi2wwe}
\chi^2_{\rm x}= \sum_{i=1}^{n^*} {\frac{{ (T_{\rm proj,i}({\bf q})-T^*_{\rm proj,i})}^2 }
{\sigma^2_{T^*_{\rm proj,i}}  }}\
\end{equation}
being $T^*_{\rm proj,i}$ the observed projected temperature profile in the $i$th ring and $T_{\rm proj,i}({\bf q})$ the convenient projection \citep[following][]{mazzotta2004} of the theoretical 3D temperature $T({\bf q})$ recovered by solving the hydrostatic equilibrium equation, once we inferred the gas density $n_e(r; {\epsilon_{c'}})$ from the brightness and we assume a gNFW parametrization for the DM $\rho_{\rm DM}=\rho_{\rm DM}({\bf {r, q}})$. ${\mathcal{L}}_{\rm lens}$ reads:
\begin{equation}\label{aa2w2q}
\!{\mathcal{L}}_{\rm lens}=\frac{\exp \left\{-\frac{1}{2}{[( k({\bf q})-k^*)]}^{\rm t}\mathbfit{C}^{-1} [( k({\bf q})-k^*)]\right\}} {(2\pi)^{m^*/2}|\mathbfit{C}|^{1/2}},
\end{equation}
where $\mathbfit{C}$ is the covariance matrix referred to the projected mass profile from lensing data including systematic effects (see below), $|\mathbfit{C}|$ indicates the determinant of $\mathbfit{C}$, $k^*=(k_1^*,k_2^*,...,k_{m^*}^*)$ are the observed measurements of the projected mass profile in the $m^*$ elliptical annuli, $k({\bf q})$ the theoretical projected mass profile retrieved by our triaxial DM model.

For the covariance matrix $\mathbfit{C}$, the below expression holds:
\begin{equation}\label{aartb}
\mathbfit{C}={\mathbfit{C}}'\; + \sigma^2_{\rm sys}\; {\mathcal{I}}
\end{equation}
being ${\mathbfit{C}}'$ the covariance matrix among the lensing measurements, ${\mathcal{I}}$ is the identity matrix and $\sigma_{\rm sys}$ a bias parameter estimator arising from measurements of systematics involved in the SL analysis. In order to calculate $\sigma_{\rm sys}$ we assumed that systematic errors can be described as gaussian errors via a diagonal matrix $\sigma^2_{\rm sys}\; {\mathcal{I}}$ with the same value in each of the diagonal elements. We checked that this simplified assumption does not affect significantly the average value of the physical parameters, while it slightly increases (10-20\%) their errors.

We marginalized over $({\bf q}, \sigma_{\rm sys})$ and therefore we have ${\mathcal{L}}={\mathcal{L}}({\bf q}, \sigma_{\rm sys})$.

\begin{figure}[!ht]
\begin{center}
\epsscale{1.05}
\plotone{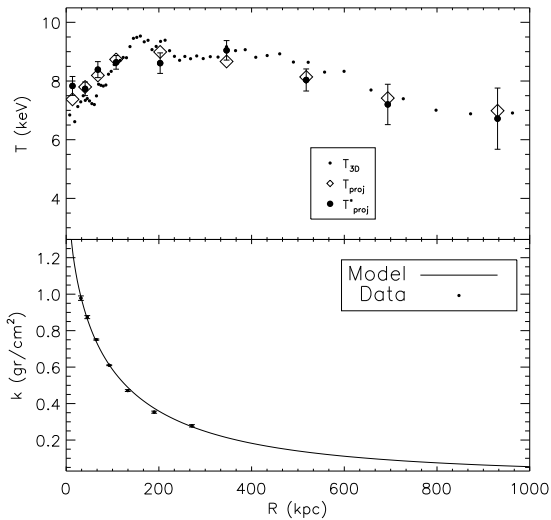}
\caption[]{Joint X-ray and lensing analysis for Abell~1689. In the upper panel we display the two quantities which enter in the X-ray analysis spectral deprojection analysis (eqn. \ref{chi2wwe}): the observed spectral projected temperature $T^*_{\rm proj,i}$ (big points with errorbars) and the theoretical projected temperature $T_{\rm proj,i}({\bf q})$ (diamonds). We also show the theoretical 3D temperature $T_j({\bf q})$ (points), which generates $T_{\rm proj,i}({\bf q})$ through convenient projection techniques. In the lower panel we display the two quantities which enter in the lensing analysis (eqn. \ref{aa2w2q}): the observed surface mass profile $k_i^*$ (points with errorbars) and the theoretical one $k({\bf q})$ (solid line). The distance of the points for both $T_j({\bf q})$ and $k_i^*$ is representative of the true spatial resolution of X-ray and lensing data, respectively.}
\label{entps332333}
\end{center}
\end{figure}

So we can determine the physical parameter of the cluster, for example the 3D temperature $T_j=T_j({\bf q})$ in the $j$th shell and the elongation $\epsilon_{c'}(e_{c'})$ of the ICM(DM) along the line of sight, just by relying on the hydrostatic equilibrium equation and on robust results of the hydrodynamical simulations of the DM profiles. In Fig. \ref{entps332333} we present an example of a joint analysis for Abell~1689: notice that in the joint analysis both X-ray and lensing data are very well fitted by our model, with a total $\chi^2_{\rm tot}=\chi^2_{\rm x}+\chi^2_{\rm lens}=7.4$ (11 degrees of freedom), $\chi^2_{\rm x}=5.5$ (5 degrees of freedom) and $\chi^2_{\rm lens}=1.9$ (2 degrees of freedom), with $\chi^2_{\rm lens}\propto -2\log({\mathcal{L}}_{\rm lens})$.

\section{Results}\label{dataan2}
\begin{figure}[!ht]
\begin{center}
\epsscale{1.05}
\plotone{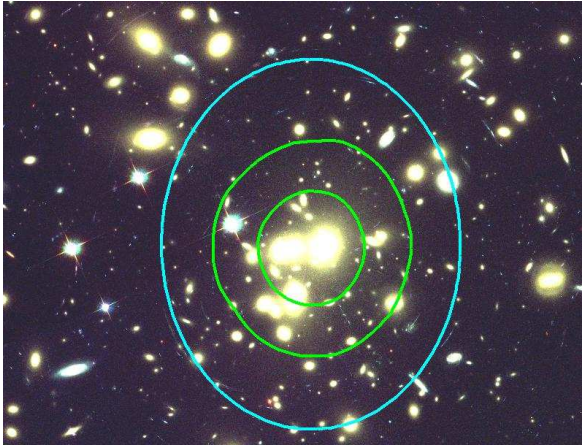}
\caption[]{Optical image of the core of Abell~1689 from the NASA/ESA Hubble Space Telescope, with overplotted the projected total mass contours computed from the gravitational lensing analysis (blue line) and from the X-ray surface brightness (green line). The north-eastern sector has been masked out in our joint SL+X-ray analysis.}
\label{entps3xc}
\end{center}
\end{figure}

In table \ref{tabdon} we present the best model fit parameters for our analysis of Abell~1689. In Fig. \ref{entps3xc} we present an image of the core of Abell~1689 from optical (Hubble Space Telescope) observations, with overplotted the projected total mass contours computed from the gravitational lensing analysis (blue line) and from the X-ray surface brightness (green line). Our findings outline a picture where Abell~1689 is a triaxial galaxy cluster with DM halo axial ratios $\eta'_{\rm{DM},b}=1.24\pm 0.13$ and $\eta'_{\rm{DM},c}=2.37\pm 0.11$, where $\eta'_{\rm{DM},b}$ is the axial ratio of the DM on the plane of the sky inferred from lensing measurements, and $\eta'_{\rm{DM},c}$ the axial ratio of the DM along the line of sight inferred through our joint analysis (see table \ref{tabdon}). Notice these elongations are statistically significant, i.e. it is possible to disprove the spherical geometry assumption.

The axial ratio of the gas is $\eta_{\rm{gas},b'}\sim 1.1-1.06$ (on the plane of the sky) and $\eta_{\rm{gas},c'}\sim 1.6-1.3$ (along the line of sight), moving from the center toward the X-ray boundary.

Here we focus on the implications of our method on the CDM scenario and on the discrepancy between X-ray and lensing masses on Abell~1689, showing that this is dispelled if we account explicitly for a 3D geometry.

\subsection{Probing the CDM scenario}\label{conclusion33a}
Measuring the three-dimensional mass distribution on galaxy cluster scales is a crucial test of the CDM scenario for structure formation models, providing constraints on the nature of dark matter. Recent works investigating mass distributions of individual galaxy clusters (e.g. Abell~1689) based on gravitational lensing analysis have shown potential inconsistencies between the predictions of the CDM scenario relating halo mass to concentration parameter, and the relationships as measured in massive clusters.

For example \cite{Broadhurst2008} using the distribution of halo profiles from \cite{neto2007} found that the predicted Einstein radii under the assumption of spherical geometry are a factor of two below the observed Einstein radii of four massive clusters with spectacular Einstein rings (among them Abell~1689). Relatively high concentrations parameters of $\sim 8-14$ are derived from lensing analysis of Abell~1689 \citep{broadhurst2005,Limousin2007}. These values are larger than the concentration parameter expected based on simulations of the standard CDM model \citep[$c\sim 4$,][]{neto2007}. Given that the predicted mass profile is too shallow compared to the observed ones, the question arises whether the projected critical surface density for lensing can be exceeded within a substantial radius for this model.

However, we emphasize the previous analyses employ standard spherical modeling of the DM halo, while numerical simulations predicts that DM halos show axis ratios typically of the order of $\sim 0.7$ \citep{Shaw2006}, disproving the spherical geometry assumption. \cite{morandi2010a} demonstrated that the halo triaxiality could cause a significant bias in estimating the desired physical parameters, i.e. concentration parameter $c$, inner slope of the DM $\alpha$ and total mass if a spherical halo model is a priori assumed for the model fitting. As a consequence, the projected mass distributions of the clusters have larger concentration parameter and inner slope of the DM compared with typical clusters with similar redshifts and masses.

In light of the previous considerations, we evaluated the Einstein radius for Abell~1689 via our triaxial joint X-ray+SL analysis. The Einstein radius $\theta_{\rm E}$ occurs at a projected radius where the mean enclosed convergence is equal to 1. Our triaxial joint X-ray+SL analysis predicts $\theta_{\rm E}=42.7\pm3.1$ arcsec for $z_s=1$, which is in agreement with the observed value of $45$ arcsec \citep[][]{broadhurst2005,Limousin2007,Richard2009}. We conclude that the large Einstein radius observed in Abell~1689 is not in conflict with CDM predictions, as long as the triaxiality of the DM halos are taken into account. In this perspective, we also find that the minor-major principle axis ratio $\eta'_{\rm{DM},c}=2.37\pm 0.11$ is consistent with the results from numerical simulations within $\sim 2.5 \,\sigma$ \citep{Shaw2006}. 

Then we focus on the determination of the other parameters of the DM halos, namely the concentration parameter and the inner slope of the DM. In this perspective, one of the main result of the presented work is to measure a central slope of the DM $\alpha=0.90\pm 0.05$ by accounting explicitly for the 3D structure for Abell~1689. This value is close to the CDM predictions of \cite{navarro1997} (i.e. $\alpha=1$), but it is even in better agreement with the more recent numeral simulations of \cite{Merritt2006}, which predicts slightly shallower inner slope. The value of the concentration parameter is $4.58\pm 0.34$, in agreement with the theoretical expectation from hydrodynamical simulations of \cite{neto2007}, where $c\sim 4$ at the redshift and for the virial mass of Abell~1689, and with an intrinsic scatter of $\sim 20$ per cent. This lends support to our insights about the role of the effects of geometry on the physical properties and allows to solve the arisen potential inconsistencies between the predictions of the CDM scenario and the measurements in massive clusters. 

If we carry out a standard spherical modeling, we obtain the biased value $\alpha =1.16\pm 0.04$ for an X-ray-only analysis, value larger than that in table \ref{tabdon}. The different value of $\alpha$ in triaxial and spherical case shows that the systematics involved in neglecting elongation/flattening of the sources along the line of sight are relevant: this likely justifies the large scatter of $\alpha$ found in the literature \citep{ettori2002b,Gavazzi2005,Sand2008,Bradac2008,Limousin2008,Biviano2006}.

\begin{table*}[!ht]
\begin{center}
\caption{Best model fit parameters of Abell~1689. The columns $1-5$ refer to the best fit parameters $c,r_{\rm s},\alpha, \eta_{DM,c}$ and $\sigma_{\rm sys}$, while the last three columns refer to the mass and radius at $\Delta=2500$, respectively, and to the Einstein radius}
\begin{tabular}{c@{\hspace{.7em}} c@{\hspace{.7em}} c@{\hspace{.7em}} c@{\hspace{.7em}} c@{\hspace{.7em}} c@{\hspace{.7em}}  c@{\hspace{.7em}}  c@{\hspace{.7em}}}
\hline \\
 $c$ & $r_{\rm s}$  & $\alpha$ & $\eta_{DM,c}$ & $\sigma_{\rm sys}$ & $M_{2500}$ & $R_{2500}$ & $\theta_{\rm E}(z_s=1)$ \\
     &     (kpc)    &           &              &${(\rm gr /cm^2)}$ & ($10^{14}M_{\odot}$) & (kpc) & (arcsec)\\
\hline \\
  $4.58\pm 0.34$ & $445\pm 35$  &  $0.90\pm 0.05$  &  $2.37\pm 0.11$ & $0.004 \pm 0.002$ & $8.58 \pm 0.23$  & $556\pm 12$ & $42.7\pm3.1$\\
\hline \\\\
\end{tabular}
\label{tabdon}
\end{center}
\end{table*}

\subsection{Resolving the discrepancy between X-ray and strong lensing masses}\label{conclusion33bbv}
Here we briefly summarize the major findings in the literature for Abell~1689 in order to study the discrepancy between the mass determined from X-ray and gravitational lensing observations.

A recent joint Chandra, HST/ACS, and Subaru/Suprime cam analysis by \cite{lemze2008} suggested that the temperature of A1689 could be as high as $T = 18$ keV at 100 $h^{-1}\, {\rm kpc}$, almost twice as large as the observed value at that radius. The derived 3D temperature profile was based on the X-ray surface brightness, the lensing shear, and the assumption of hydrostatic equilibrium. From the disagreement between the observed X-ray temperature and the deduced one, they concluded that denser, colder, and more luminous small-scale structures could bias the X-ray temperature. Nevertheless \cite{Peng2009} proved that if the temperature profile of the ambient cluster gas is in fact that of \cite{lemze2008}, the cool clumps would have to occupy 70-90 per cent of the space within 250 kpc radius, assuming that the two temperature phases are in pressure equilibrium. They conclude that the scenario proposed by \cite{lemze2008} is unlikely.

Since lensing is sensitive to the integrated mass contrast along the line of sight, either fortuitous alignments with mass concentrations which are not physically related to the galaxy cluster or departures of the DM halo from spherical symmetry can justify the discrepancy in the literature between cluster masses determined from X-ray and strong gravitational lensing observations, the latter being significantly higher than the former \citep{Gavazzi2005}. 

\cite{lokas2006} pointed out that Abell~1689 has a complex structure in velocity space, suggesting the presence of dynamically independent structures along the line of sight, which would affect lensing mass estimates, but \cite{lemze2009} disagree with this projection view and they argued that there is no evidence for such substructure in their velocity data. They conclude that only one identifiable substructure at $+3000$ km/s, $1.50$ arcmin to the NE, which is seen in the strong lensing mass analysis but is determined not to be massive (less than 10 per cent of the total mass in the strong lensing region). Nonetheless, the higher than usual velocity dispersion in the cluster center, $\sim 2100$ km/s, indicates that the central part is quite complex \citep{Czoske2004}. This may also imply that the halo is elongated in the line-of-sight direction, as galaxies move faster along the major axis. 

When it comes to a "superlens" clusters as Abell~1689, halo sphericity is never a justified assumption. Indeed \cite{Oguri2009} showed that SL clusters with the largest Einstein radii constitute a highly biased population with major axes preferentially aligned with the line of sight increasing the magnitude of lensing, and \cite{Oguri2005} concluded that Abell~1689 weak lensing measurements are indeed compatible with the CDM-based triaxial halo model if Abell~1689 represents a rare population ($\sim$6 per cent fraction) of cluster-scale halos. \cite{morandi2010a} demonstrated that triaxiality allow to solve the mass discrepancy between the lensing and X-ray estimates in the galaxy cluster MACS\,J1423.8+2404. 

\begin{figure}[!ht]
\begin{center}
\epsscale{1.05}
\plotone{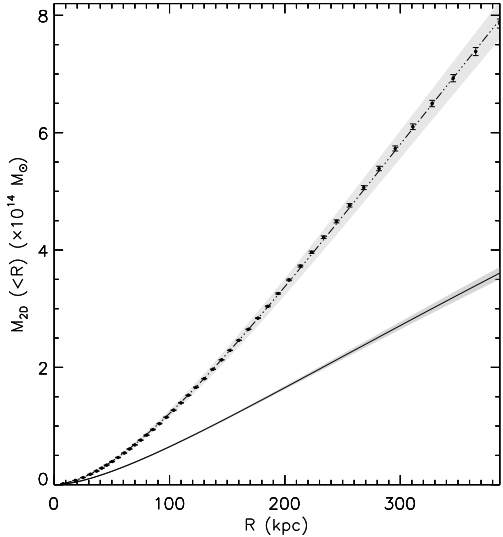}
\caption[]{2D masses enclosed within a circular aperture of radius $R$ from lensing data (points with errorbars), from an X-ray-only analysis under the assumption of spherical geometry (solid line with the 1-$\sigma$ error gray shaded region), and from a joint X-ray+lensing analysis taking into account the 3D geometry (dot-dashed line with the 1-$\sigma$ error gray shaded region).}
\label{entpsr3}
\end{center}
\end{figure}

Indeed, in Fig. \ref{entpsr3} we compare the 2D mass enclosed within circular apertures of radius $R$ for lensing, for an X-ray-only analysis under the assumption of spherical geometry, and for a joint X-ray+lensing analysis taking into account the 3D geometry. We emphasize the good agreement between the masses inferred from lensing and a joint analysis based on triaxial modeling. On the contrary an X-ray-only analysis based on the standard spherical modeling clearly predicts systematically lower masses by $\sim$ a factor of two in the radial range between 30 and 400 kpc.

This confirms our insights about the role of the effects of geometry on the physical properties and solve the long-standing discrepancy between X-ray and strong lensing mass of Abell~1689.

\section{Summary and conclusions}\label{conclusion33}
In this paper we have employed a triaxial halo model for the galaxy cluster Abell~1689 to extract more reliable information on the three-dimensional shape and physical parameters, by combining X-ray and strong lensing measurements. 

We demonstrated that the halo triaxiality could cause a significant bias in estimating the desired physical parameters, i.e. concentration parameter $c$, inner slope of the DM $\alpha$ and total mass if a spherical halo model is a priori assumed for the model fitting.

We focused on the implications of our method on the CDM scenario, proving that the value of the $c$ and $\alpha$ are in agreement with the CDM predictions, once we properly accounted for the 3D shape of the cluster. Departures of $c$ and $\alpha$ from the theoretical expectation of the CDM scenario found in the literature can be explained by a halos having the major axis preferentially oriented toward the line of sight. In particular, accounting for the 3D geometry allows to resolve the long-standing discrepancy between X-ray and strong lensing mass  of Abell~1689 in literature and predicts an Einstein radius in agreement with the observations.

\acknowledgments
The Dark Cosmology Centre is funded by the Danish National Research Foundation. K.P. acknowledges support from Instrument Center for Danish Astrophysics. M.L. acknowledges the Centre National d'Etude Spatiale (CNES) and the Centre National de la Recherche Scientifique (CNRS) for its support. M.L. acknowledges support from the city of Marseille through an installation grant. We thank Johan Richard for his help with Fig. \ref{entps3xc} and useful discussion.

{\it Facilities:} \facility{HST}, \facility{CXO}


\end{document}